\documentstyle[twoside,fleqn,espcrc2]{article}

\newcommand{\AmS}{{\protect\the\textfont2
 A\kern-.1667em\lower.5ex\hbox{M}\kern-.125emS}}
        \newcommand{\be}{\begin{equation}}
        \newcommand{\ee}{\end{equation}}
        \newcommand{\bea}{\begin{eqnarray}}
        \newcommand{\eea}{\end{eqnarray}}
        \newcommand{\za}{< \!\!  }
        \newcommand{\ze}{  \!\! >}
        \def \3{\ss}

\title{\rightline{DESY 94-208}  \rightline{hep-th/9411094}
   \bigskip On the Local Equilibrium Condition}

\author{ Hermann He{\ss}ling
        \\
        Deutsches Elektronen--Synchrotron (DESY),
        Notkestra\3e 85, D--22603 Hamburg, Germany
       }

\begin{document}

\begin{abstract}

A  physical system is in local equilibrium if it cannot be
distinguished from a global equilibrium by ``infinitesimally localized
measurements''.
This should be a natural  characterization of local equilibrium, but
the problem is to give a precise meaning to the qualitative phrase
``infinitesimally localized measurements''.
A solution is suggested  in form of a
{\em Local Equilibrium Condition} (LEC), which
can be applied to linear relativistic quantum
field theories but  not directly to selfinteracting quantum fields.
The concept of {\em local temperature}  resulting from LEC
is compared to an old
approach to local temperature  based  on  the
principle of maximal entropy.
It is shown that the principle of maximal entropy does
not always  lead
to physical states if it is applied to relativistic quantum field
theories.

\end{abstract}

\maketitle

\section{Introduction}

An understanding of nature seems to be easier
at very small length scales
than at larger length scales. This is our every day experience
here at DESY, where we analyze ZEUS data
\footnote{
  For a description of the ZEUS detector
  see e.g. \cite{zeus}.}
to test
predictions of
the asymptotically free QCD.

The smaller the observables the less can be resolved:
whatever the state of a physical system is, it
cannot be distinguished from the vacuum state if the
localization regions of the observables are shrinked to a point.
This is the content
of the Principle of Local Stability \cite{hns}, \cite{haag}.
What can be said about the state of a physical system
if the localization region of a measurement is not
completely  shrinked to a point, but is ``infinitesimally localized''?
 From general relativity  we know that because of the
Equivalence Principle  gravitation
is locally constant.
In \cite{hessling} a formulation of a Quantum Equivalence Principle (QEP)
was suggested.
According to QEP
the states of all physical systems
are locally constant.
QEP was investigated in
the Rindler spacetime
\footnote{The Rindler spacetime is a wedge in the
    Minkowski spacetime ($ | t| < x^{(1)}$).
    It is  a simple model of a black hole.}
and it was shown
  that the Hawking--Bisognano--Wichman
temperature \cite{hawking}, \cite{bw1}, \cite{bw2} is
 a consequence of  QEP \cite{hessling}.

In this paper we try
to characterize local equilibrium states
by formulating a condition
for
``infinitesimally localized measurements''.
This condition is presented in the next section. The next section
starts
with a collection of some
 more or less known facts and a presentation of
our nomenclature. In the last section we compare
our approach to local equilibrium to an approach based
on the principle of maximal entropy.

\section{Local Equilibrium Condition}

In a quantum mechanical system of finite degrees of freedom the expectation
value $\za A \ze$ of any observable $A$
in a state $\za \cdot \ze$  can be characterized by a
density matrix $\rho$
\bea
       \za A \ze &=& \frac{
                {\rm   Tr}\, \rho A}{ {\rm Tr}\, \rho}
  .     \label{poltamann-mod}
\eea
If one introduces  the
{\em modular Hamiltonian} $\tilde H$ by
\bea
    e^{- {\tilde \beta} {\tilde {H}} } &=& \rho
       \nonumber
\eea
where $\tilde \beta$ is a number introduced for later convenience,
the {\em modular
evolution}
\bea
  \gamma_\tau (A) = e^{i{\tilde {H}}\tau}Ae^{-i{\tilde {H}}\tau}
  \nonumber
\eea
can be defined. Cyclicity of the trace gives
the  {\em KMS--condition} \cite{kubo}, \cite{ms}
\bea
        \za \gamma_\tau(A) B \ze
         &=& \za B \gamma_{\tau + i {\tilde \beta}}(A) \ze
         \label{kms-mod}.
\eea
In quantum field theory the right hand side of
 (\ref{poltamann-mod})
does not
exist but the
KMS--condition  (\ref{kms-mod})
can be used directly to characterize the state \cite{hhw}, \cite{haag}.
For simplicity we   concentrate on
the Klein--Gordon field $\phi(x)$ in Minkowski spacetime.

If the system is in a global equilibrium
state $\za \cdot \ze_\beta$  with temperature $1/\beta$ the
modular Hamiltionan is nothing but  the  Hamilton operator,
$\tilde H = H$, which generates
the time evolution $\alpha_t$ along an inertial time
coordinate $t$  in the Minkowski
spacetime
\bea
         \alpha_t \phi(0,{\vec x}) \;=\;
   e^{iHt} \phi(0,{\vec x}) e^{-iHt} &=&
  \phi(t,{\vec x})
 .   \nonumber
\eea
The parameter $\tilde \beta$ represents  the inverse  temperature
$\beta$ of the system.
If we replace the modular evolution $\gamma_\tau$ in  (\ref{kms-mod})
by the time evolution $\alpha_t$ and
perform a  Fourier transformation of the
KMS--condition, the 2--point function $\za \phi(x') \phi(x) \ze_\beta$
is representable in terms of a state independent  commutator
\cite{hessling}
\bea
  \za \phi(x') \phi(x) \ze_\beta \;=
\;\;\;\;\;\;\;\;\;\;\;\;\;
\;\;\;\;\;\;\;\;\;\;\;\;\;\;
\;\;\;\;\;\;\;\;\;\;\;\;\;\;
   \\ \nonumber
\;\;\;\;\;
   \frac{i}{2\tilde \beta}
   \int {\rm d} \tau [ \alpha_\tau \phi(x),\phi(x')]
   \coth \frac{\pi}{\tilde \beta}(\tau - i \epsilon)
\label{kmskomm}
\eea
Using  the
well known non--equal time
 commutator for a massless Klein--Gordon
field
\bea
   [\phi(x),\phi(x')] = \frac{1}{2\pi i} {\rm sign}(t-t')
    \delta( (x-x')^2)
\nonumber
\eea
we obtain for the 2--point function
\bea
  \za \phi(x') \phi(x) \ze_\beta =
  \frac{1}{4 \pi \beta {\dot \sigma}}
 \left( \coth \frac{\pi \tau_+}{\beta} -\coth \frac{\pi \tau_-}{\beta}
 \right)
 \label{result}
\eea
where
\bea
    \tau_\pm = t' - t \pm \sqrt{ ({\vec x}'- \vec x)^2 } - i \epsilon,
\;\;
   {\dot \sigma} = 2 \sqrt{ ({\vec x}'- \vec x)^2 }
\nonumber
\eea

We understand
local equilibrium  as a state
which  cannot be distinguished from
a global equilibrium state by ``infinitesimally localized measurements''.
If the localization region of an  observable $A$ is made smaller and
smaller
the expectation value $\za A \ze$ of the observable $A$ in a
local equilibrium state $\za \cdot \ze$
should become
more and more identical to the expectation value $\za A \ze_{\beta_*}$
of
$A$ in an equilibrium state $\za \cdot \ze_{\beta_*}$
at a certain temperature $\beta_*$.
To describe the shrinking of the localization region of an observable $A$
 we use a one--parametric scaling procedure $\delta_\lambda A$.

For the $n$--point observable
$A=\phi(x_1) \dots \phi(x_n)$ the scaling procedure is defined as
\cite{hns}, \cite{fh}, \cite{hessling}
\bea
      \delta_\lambda  A
      &=&
      N(\lambda)^n \; \phi(\chi_\lambda x_1)\dots \phi(\chi_\lambda x_n)
   \label{scalen}
\eea
where (with respect to inertial coordinates $x^\mu$)
\bea
    (\chi_\lambda x)^\mu ={ x_*}^\mu + \lambda
                                   ( x^\mu - {x_*}^\mu)
     \label{lammel}
\eea
is a 1-parametric scaling
                  diffeomorphism with        $\chi_1 x = x$
and $\chi_0 x = x_*$.
In the limit $\lambda \rightarrow 0$ the localization points $x_1
\dots x_n$
of the  $n$--point observable $\delta_\lambda A$ are
scaled into the point
$x_*$.
The {\em  scaling function} $N(\lambda)$
has to be adjusted in such a way
that the
{\em scaling limit}  of the $n$--point function,
$
  \lim_{\lambda \rightarrow 0} \za \delta_\lambda A  \ze,
$
is well defined.
  A suitable scaling function for the Klein--Gordon field is
 $N(\lambda)= \lambda$.

According to the Quantum Equivalence Principle (QEP) \cite{hessling}
the scaled observable $\delta_\lambda A$ has to fulfil two requirements
for small values of the scaling parameter $\lambda$:
the expectation value $\za \delta_\lambda A \ze$
has to be locally  constant
around the scaling point $x_*$
and  its  the scaling limit,
$
 \lim_{\lambda \rightarrow 0} \za \delta_\lambda A \ze,
$
has to be continuous in $x_*$.
In linear  quantum  field theories
the first requirement
can be written as the extremum conditon
\bea
    \lim_{\lambda \rightarrow 0}
                                  \frac{d}{d \lambda}
                 \za \delta_\lambda A
      \ze
          &=& 0.
      \label{qap2}
\eea
Therefore
the first nontrivial information about the state of
a linear quantum field  is beyond the first
order in $\lambda$.

We say that a state $\za \cdot \ze$ fulfils the
{\em Local Equilibrium Condition} in the point $x_*$,
if  it does not differ from
a global equilibrium state $\za \cdot \ze_{\beta_*}$ of temperature
$1/\beta_*$ up to second order in the scaling parameter $\lambda$
\bea
    \lim_{\lambda \rightarrow 0}
                                  \frac{d^2}{d \lambda^2}
  \left(
       \frac{}{}
                 \za \delta_\lambda  A  \ze
      -           \za \delta_\lambda A  \ze_{\beta_*}
\right)
          &=& 0
  .    \label{lec}
\eea

For a massless Klein--Gordon field the equilibrium part of LEC can be
calculated from (\ref{result})
\bea
    \lim_{\lambda \rightarrow 0}
                                  \frac{d^2}{d \lambda^2}
                \za \lambda^2 \phi(\chi_\lambda x')
                              \phi(\chi_\lambda x)
                \ze_{\beta_*}
   &=&
    \frac{1}{12\;{\beta_*}^2}.
\label{lecsec}
\eea

Let us apply  LEC to  Hadamard states.
Hadamard states are definable
in linear quantum field theories and are
quasifree states\footnote{
            A state is called quasifree if its truncated
            $n$--point--functions vanish for $n\neq 2$.}
 with a  specific singularity structure:
the symmetric part of the   2--point--function
is identical
   with Hadamard's
                fundamental solution of the wave equation
\cite{dwb}
   \bea
     \za \{ \phi(x') , \phi(x) \} \ze
     &=&
     \frac{u}{\sigma} + v \ln \sigma + w
   \eea
where $\sigma$ is the square of the geodesic distance between $x'$ and $x$.
The functions $u, v, w$ are regular in $x$ and $x'$.
The information  about  the state is contained in $w$; $u$
and $v$ are state independent and are uniquely fixed by the geometry of
the spacetime.
It is conjectered that the 2--point function of any
physical state of the Klein--Gordon field can locally be approximated
by a Hadamard state
 \cite{haag}. Assuming the validity of this conjecture
\footnote{
   For a massless Klein--Gordon field the following conclusions
   can directly be proven without this conjecture
   by using (\ref{result}).
    }
the equilibrium state can be approximated by a state
of the form
\bea
     \za \{ \phi(x') , \phi(x) \} \ze_{\beta_*}
     &=&
     \frac{u}{\sigma} + v \ln \sigma + w_{\beta_*}.
\eea
It follows that LEC reduces to
\bea
          w(x_*,x_*) = w_{\beta_*}(x_*,x_*)
  \label{wwbe}
\eea
since the state independent singular parts $u/\sigma,v \ln \sigma$ cancel
because of the difference in
(\ref{lec})
and since the state dependent parts $w(\chi_\lambda x',\chi_\lambda x)$ and
$
  w_{\beta_*}(\chi_\lambda x',\chi_\lambda x)
$ are regular in the limit $\lambda \rightarrow 0$.
This means
that LEC does not depend on the scaling function
$\chi_\lambda$ in the sense that  Eqn (\ref{lammel}) can be
replaced by any one--parametric scaling
diffeomorphism $\chi_\lambda$ which has $x_*$ as a
fixpoint. We are therefore allowed to call $1/\beta_*$
a {\em local temperature}.

Combining (\ref{lecsec}) and (\ref{wwbe}) we see
that
 all Hadamard states of the  massless Klein--Gordon field
with a non--negative $w(x_*,x_*)$ have the local temperature
\bea
    1/\beta_* = \sqrt{12 \; w(x_*,x_*)}
\nonumber
\eea
in the scaling point $x_*$.
Hadamard states with a negative $w(x_*,x_*)$  have
an imaginary local temperature and therefore cannot be
local equilibrium states.

How can one describe local equilibrium states in curved spacetimes?
In curved spacetimes the
existence of global equilibrium states
can no longer  be expected. Therefore LEC cannot be
applied directly.
Nevertheless let us insist on our ``global first -- local next'' point of view:
before one can define local equilibrium one has to know what
equilibrium in a  {\em finite} region of spacetime is.
If a physical system is influenced by a rapidly
 changing gravitational force, it is
intuitively clear that
 the system has  to react, i.e. it has  to change its state.
But stationarity is an important characteristic quality of equilibrium.
We therefore exclude non--stationary gravitational fields
from the
finite spacetime region $\cal O$
where we would like to investigate local equilibrium states.
This can be done by assuming that in the region $\cal O$
there is a  timelike
Killing field $\partial/\partial t$.
Quantitatively
we characterize equilibrium states in the region $\cal O$
by carrying over the concept of
{\em local KMS--states} from the Minkowski spacetime
\cite{bj}.
A local KMS--state $\za \cdot \ze_{\beta, {\cal O}}$
in the region $\cal O$ of temperature $1/\beta$
has a 2--point function
$
  \za \phi(x') \phi(t,{\vec x}) \ze_{\beta, {\cal O}}
$
which is analytical in the open interval
$0< {\rm Im}\, t < \beta$, continuous in the closed interval
$0 \le {\rm Im}\, t \le \beta$
and fulfils the KMS--conditon
\bea
        \za \phi(x) \phi(x') \ze_{\beta, {\cal O}}
       &=&
        \za \phi(x')\phi(t+i\beta,{\vec x}) \ze_{\beta, {\cal O}}
\nonumber
\eea
for all points $x=(t,{\vec x}), x'$ in $\cal O$.
(The step from the  $2$--point function
to arbitrary observables  is straightforward and therefore not writen down.)
To formulate LEC in the spacetime region $\cal O$, one  has
to replace the reference state $\za \cdot \ze_{\beta_*}$ in
(\ref{lec}) by the local KMS--state $\za \cdot \ze_{\beta_*, {\cal O}}$.

In \cite{hessling} it was shown that the derivative condition
(\ref{qap2}) has to be modified if one wants to
formulate QEP for  asymptotically free quantum field theories
like QCD, since
the running coupling constant does not
smoothly become zero in the short distance limit $\lambda \rightarrow
0$, but logarithmically. Because of the same reason
 the second derivative condition
(\ref{lec}) needs a modificaton if one wants to characterize local
equilibrium states
for selfinteracting quantum fields.


\section{LEC versus the principle of maximal entropy}

Global equilibrium states
in nonrelativistic quantum theories
 can be obtained by the  extremalization of
a certain functional: the entropy.
The entropy of a state $\za \cdot \ze = {\rm Tr \hat \rho (\cdot)}$
is defined as
\bea
       S = - {\rm Tr} {\hat \rho} \ln {\hat \rho}
       \nonumber
\eea
where $\hat \rho = \rho/{\rm Tr} \rho$ is the normalized density
matrix.
Consider the set of states with a given  expectation value
for the Hamilton operator
\bea
         E = \za H \ze
   \label{enerconstr}
\eea
A global equilibrium state is characterizeable as  the state
of
 maximal entropy within this set.
 From the
variational equation for the normalized density
matrix,
$
 \delta ( S + c(1 - \za 1 \ze)
 + \beta(E- \za H\ze) )
$
$
 = 0,
$
where $c$ and $\beta$ are  Lagrange multipliers for
the normalization condition ${\rm Tr}\hat \rho = 1$ and the constraint
(\ref{enerconstr}) respectively, one gets
\bea
        \rho = e^{-\beta H}
\nonumber
\eea
It turns out that
$\beta$
is the inverse temperature of the system.

To determine the state of a system with a nonuniform
temperature it
was suggested \cite{zubarev} that the principle of
maximal entropy has to be applied to a local form of the constraint
(\ref{enerconstr})
\bea
          \epsilon (\vec x) = \za {\cal H}(\vec x) \ze
\label{locconstr}
\eea
where ${\cal H} (\vec x)$ is the Hamiltion density at time zero.
This leads to
 a continuum of Lagrange multipliers $\beta(\vec x)$ and
the density matrix
\bea
           \rho = e^{-\int {\rm d}^3 x \beta(\vec x) {\cal H} (\vec x)}
 . \label{locden}
\eea
$1/\beta(\vec x)$
is interpreted as the {\em local temperature} of the
system \cite{zubarev}.
For the current status of this approach we refer to
\cite{zubarev2} and references therein.

What is the relation between the local temperature $1/\beta_*$
introduced in the last section and the local temperature $1/\beta(\vec x)$
obtained with the principle of maximal entropy?

Let us choose a $\beta({\vec x})$  which is linear in
  $x^{(1)}$
\bea
      \beta(\vec x) = \tilde \beta x^{(1)}
\nonumber
\eea
Then the  modular Hamiltonian of the density matrix (\ref{locden})
\bea
     \tilde H = \int {\rm d}^3 x\; x^{(1)} {\cal H} (\vec x)
\label{htild}
\eea
becomes the generator of a boost transformation in
$x^{(1)}$--direction
and for  the modular evolution of the Klein--Gordon field one finds
\bea
        \gamma_\tau \phi(0,\vec x)
         =
 \phi(x^{(1)} \sinh  \tau, x^{(1)} \cosh  \tau,x^{(2)},x^{(3)} )
 \nonumber
\eea
This system was studied in \cite{hessling} and it was shown that
only one value of the parameter $\tilde \beta$ is allowed by QEP, namely
$\tilde \beta = 2\pi$.
We conclude that the principle of maximal entropy does not always lead
to physical states.
 Consequently the interpretation
of $1/\beta(\vec x)$ as a local temperature does not make
sense in general.

For the allowed parameter $\tilde \beta = 2 \pi$ the 2--point function
of the Klein--Gordon field in state given by (\ref{htild}) is just
the 2--point function of the vacuum state in the
Minkowski spacetime \cite{bw1}, \cite{bw2}, i.e.
the local temperature in the sense of LEC is identically zero
\bea
           1/\beta_* =0
\nonumber
\eea
On the other hand the local temperature from the principle of maximal entropy
\bea
      \frac{1}{\beta(\vec x)} = \frac{1}{2\pi x^{(1)}}
\nonumber
\eea
is the Unruh temperature of an uniformly accelerated detector, where
the Unruh temperature is
measured with respect to the local  proper time in the detector
\cite{haag}.

The two concepts of local temperature
refer to different time concepts.
The local temperature concept from the principle of maximal
entropy  refers to
a time defined by a certain subclass of the modular evolutions.
(It would be interesting to know under which conditions
this approach makes sense physically.)
In the Minkowski spacetime
the local temperature concept of LEC is
 related to
a
time given by the clock of
an experiment at rest.

\section*{Acknowledgments}

We would like to thank R. Haag for helpful
and stimulating discussions. We also thank U. Bannier,
                                          D. Buchholz,
K. Fredenhagen,    K.--H. Rehren   and R. D. Tscheuschner
                                for discussions.
The support from the ZEUS collaboration is acknowledged
with great appreciation.

\end{document}